\input harvmac.tex
\def\np#1#2#3{Nucl. Phys. {\bf B#1} (#2) #3}
\def\pl#1#2#3{Phys. Lett. {\bf #1B} (#2) #3}
\def\prl#1#2#3{Phys. Rev. Lett.{\bf #1} (#2) #3}
\def\physrev#1#2#3{Phys. Rev. {\bf D#1} (#2) #3}

\noblackbox

\Title{RU-97-8, hep-th/9703052}
{\vbox{\centerline{Zero and One-dimensional Probes}
        \vskip4pt\centerline{with N=8 Supersymmetry}}}
\centerline{Tom Banks, Nathan Seiberg, and Eva Silverstein 
\footnote{$^\star$}
{banks, seiberg, evas@physics.rutgers.edu 
}}
\bigskip\centerline{Department of Physics and Astronomy}
\centerline{Rutgers University}
\centerline{Piscataway, NJ 08855-0849}

\vskip .3in

We study the symmetry structure of $N=8$ quantum mechanics, and apply it
to the physics of D0-brane probes in type I' string theory.  We focus on
the theory with a global $Spin(8)$ R symmetry which arises upon dimensional
reduction from $2d$ field theory with $(0,8)$ supersymmetry.  There are
several puzzles involving supersymmetry which we resolve.  In
particular, by taking into account the gauge constraint and central
charge we explain how the system preserves supersymmetry despite having
different numbers of bosonic and fermionic variables.  The resulting
zero-point energy leads to a linear potential consistent with
supersymmetry, and the metric is largely unconstrained.  We discuss
implications for type I' string theory and the matrix model proposal for
M theory.

\Date{2/97} 

\newsec{Introduction}

\nref\probes{M. Douglas and G. Moore, ``D-branes, Quivers, and ALE
Instantons'', hep-th/9603167.}%
\nref\doug{M. Douglas, ``Gauge Fields and D-branes'', hep-th/9604198.}%

\nref\bds{T. Banks, M. Douglas, and N. Seiberg, \pl{387}{1986}{278},
hep-th/9605199.}%
\nref\seithree{N. Seiberg, \pl{384}{1996}{81}, hep-th/9606017.}%
\nref\dkps{M. Douglas, D. Kabat, P. Pouliot, and S. Shenker,
``D-branes and Short Distances in String Theory'', hep-th/9608024.}%
\nref\seiberg{N. Seiberg, ``Five-Dimensional SUSY Field Theories,
Non-trivial Fixed Points and String Dynamics'', hep-th/9608111.}%
\nref\morsei{D. Morrison and N. Seiberg, ``Extremal Transitions and
Five-Dimensional Supersymmetric Field Theories'', hep-th/9609070.}%
\nref\aks{O. Aharony, S. Kachru, and E. Silverstein, ``New $N=1$ 
Superconformal Field Theories in Four Dimensions from D-brane Probes'',
to appear in {\it Nucl. Phys. B}, hep-th/9610205.}%
\nref\asy{O. Aharony, J. Sonnenschein, S. Theisen, and S. Yankielowicz,
``Field Theory Questions for String Theory Answers'', hep-th/9611222.}%
\nref\sen{A. Sen, ``F Theory and The Gimon-Polchinski Orientifold'',
hep-th/9702061.}%
\nref\dls{M. Douglas, D. Lowe, and J. Schwarz, ``Probing F Theory
with Multiple Branes'', hep-th/9612062.}%

One of the most interesting phenomena uncovered by string
duality is the relation between gauge dynamics and compactification
geometry.  A particularly simple manifestation of this correspondence
occurs in the context of brane probes.  It first appeared in
\refs{\probes, \doug} and was developed in 
\refs{\bds-\dls}.

In perturbative string theory, there is a detailed
correspondence between worldsheet physics and spacetime
physics.  In particular, worldsheet supersymmetry is
related to spacetime supersymmetry (though the two
are not identical).  Modular invariance is equivalent
to spacetime anomaly cancellation.  For large compactification
volume the target space of the worldsheet sigma model is
the compactification manifold.

It is of some interest to understand the analogues of these statements
for probes of dimension $p\ne 1$.  Many new quantum field theory results
have been obtained on brane probes using the relation to spacetime
physics.  In each case, the probe theory reflects
the geometry of spacetime in its Lagrangian (through for example the
metric appearing in the kinetic term for scalar fields).  One can
consider probes of different dimension in the same background, all of
which encode the same spacetime geometry.  In this paper we will clarify
this correspondence for $p=1$ and $p=0$ (with 8 supersymmetries in both
cases).  For $p=1$, with $(0,8)$ supersymmetry, we find that the gauge
fermions which are central to the spacetime anomaly cancellation
mechanism of the heterotic string are necessary in the D string
formalism to cancel worldsheet gauge anomalies which arise when more
than one string are present.

\nref\ulf{U. Danielsson and G. Ferretti, ``The Heterotic Life
of the D Particle'',  hep-th/9610082.}%
\nref\joenotes{J. Polchinski, S. Chaudhuri, and C. Johnson, 
``Notes on D-branes'', hep-th/9602052.}%
\nref\joenew{J. Polchinski, ``TASI Lectures on D-branes'',
hep-th/9611050.}%
\nref\ks{S. Kachru and E. Silverstein, ``On Gauge Bosons in the
Matrix Model Approach to M-theory'', to appear in {\it Phys. Lett.}
{\bf B}, hep-th/9612162.}%
\nref\motl{L. Motl, ``Quaternions and M(atrix) Theory in
Spaces with Boundaries'', hep-th/9612198.}%
\nref\lowe{D. Lowe, ``Bound States of Type I' D-particles
and Enhanced Gauge Symmetry'', hep-th/9702006.}%
\nref\rey{N. Kim and S. Rey, ``M(atrix) Theory on an
Orbifold and Twisted Membrane'', hep-th/9701139.}%

We then study the $p=0$ system that can be thought of as the dimensional
reduction of the $(0,8)$ $p=1$ system.  Aspects of this system were
studied in \refs{\ulf - \joenew} and in the matrix model context in
\refs{\ks - \rey}.  At first sight, this system has several puzzling
features.  In particular, as noted by \ulf, the number of bosonic and
fermionic quantum-mechanical variables is not equal, despite the
unbroken supersymmetry.  The deficit between the fermionic and bosonic
variables leads to a nontrivial potential for one of the fields.  In
section 3 we explain how this is consistent with the supersymmetry
algebra by keeping track of the crucial role played by the gauge
constraint.  We also find that supersymmetry imposes a very weak
constraint on the kinetic terms.  Direct computation in the type I'
system confirms the presence of a nontrivial metric even when the
dilaton tadpole cancels locally.

In the next section we study the two dimensional theory with $(0,8)$
supersymmetry and in section 3 we study its dimensional reduction to
quantum mechanics.  In section 4 we consider applications of this
quantum mechanical system to the type I' theory and in section 5 to the
Matrix model of M theory.

\newsec{The $(0,8)$ Supersymmetric Theory in $d=2$}

Let us begin by considering 1-brane probes with $(0,8)$ supersymmetry.
One example is the heterotic string (which in type I language
is the D-1-brane).  The multiplets are as follows:
There is a gauge multiplet containing a gauge boson
$A_\mu$ and 8 left-moving fermions $\lambda_{-,a}~~(a=1,\dots,8)$.
There are two types of matter multiplets.  One
type includes 8 (nonchiral) bosons $X^i,~i=1,\dots,8$ and 8 right-moving
fermions $\theta_{+,\dot a},~\dot a=1,\dots,8$.  The other type involves only
left moving fermions $\chi_-$.  

We limit the discussion to matter fields in real representations of the
gauge group, because these are the kinds of matter fields which we will
find in the applications below.  The Lagrangian
determined by supersymmetry and gauge invariance is
\eqn\Lag{\eqalign{&{\cal L}={-i\over g^2}\lambda_{-,a}{\cal D}_+\lambda_{-,a}
+{1\over g^2}Tr F_{\mu\nu}^2
+\partial_\mu X_i\partial^\mu X_i
-i\theta_{+,\dot a}{\cal D}_-\theta_{+,\dot a}\cr
&+2i\theta_{+,\dot a}\lambda_{-,a}\sigma^i_{a\dot a}X_i
-{g^2\over 4}\sum_{\alpha,i,j}(X_iT^\alpha X_j)^2
+i\bar\chi_r{\cal D}_+\chi_r-\sum_r\bar\chi_rm_r\chi_r\cr}}
Here $\alpha$ runs over the generators of the gauge group, and
$T^\alpha$ is the gauge generator in the representation
under which the $X_i, \theta_{\dot a}$ transform.  
The supersymmetry transformation laws are:
\eqn\susytwo{\eqalign{
&\delta\lambda^\alpha_a=F^\alpha_{01}\epsilon_a
-{g^2\over 4}X_iT^\alpha X_j\sigma^{ij}_{ab}\epsilon_b\cr
&\delta A_0=-i\epsilon_a\lambda_a\cr
&\delta A_1=i\epsilon_a\lambda_a\cr
&\delta\theta_{\dot a}=({\cal D}_+X_j)\sigma_{\dot a a}^j\epsilon_a\cr
&\delta X_i=i\epsilon_a\sigma_{a\dot a}^i\theta_{\dot a}\cr}}
This theory has an $Spin(8)$ global R symmetry (to be identified
with part of the spacetime Lorentz group) under which the supercharges
transform in the ${\bf 8_s}$,
$\lambda_{-,a}$ transforms in the ${\bf 8_s}$, 
$X^i$ transforms in the ${\bf 8_v}$, and $\theta_{+,\dot a}$ transforms
in the ${\bf 8_c}$.    
The fact that $\lambda_{-,a}$ and $\theta_{+,\dot a}$ transform
as spinors of opposite chirality is enforced by the
Yukawa coupling between $X^i$, $\lambda_{-,a}$ and 
$\theta_{+,\dot a}$.  

At one loop the field content is constrained by anomalies,
as also noted in this context in \rey.  
Let us denote by $r_g,r_R$, and $r_L$ the representations
of the ($\lambda_-,A_\mu$), ($X,\theta_+$) and $\chi_-$ multiplets
respectively under the gauge symmetry.  The cancellation
of the gauge anomaly requires
\eqn\anom{8C_2(r_g)+C_2(r_L)=8C_2(r_R).}      
In the case of $n$ heterotic strings, the gauge group
is $Spin(n)$.  The $(X,\theta_+)$ multiplet is in
the symmetric tensor and the 32 $\chi_-$ multiplets are in
the fundamental.   Here we are counting Majorana-Weyl fermions.
In \Lag\ these are grouped into sixteen complex fermions.
 In the $2d$ context then, anomaly
cancellation explains the presence of left-moving
fermions which lead to the spacetime $Spin(32)$ gauge symmetry.

\newsec{$N=8$ Quantum Mechanics}

The dimensional reduction of the $(0,8)$ theory considered above gives an
$N=8$ supersymmetric quantum mechanics with an $Spin(8)$ global R
symmetry.  This describes the dynamics of type I' zero branes.  By
contrast, one could also consider theories with (4,4) supersymmetry in
$2d$.  The dimensional reduction of this again has $N=8$ supersymmetry
but with global R symmetry $Spin(3)\times Spin(5) \subset Spin(8)$.  It
describes for example the dynamics of D0 branes near D4 branes in type
IIA string theory \dkps.

We will be interested in the $N=8$ supersymmetric quantum
mechanics with $Spin(8)$ global symmetry.  Let us begin
by considering a theory with only a $U(1)$ gauge multiplet.
We will later realize this as the low-energy effective
theory obtained by integrating out degrees of freedom
{}from a larger theory.  
This multiplet contains a gauge boson $A_0$, a scalar $\phi$ and eight
fermions $\lambda_a,~~a=1,\dots,8$.  Consider the Lagrangian
\eqn\coulLag{{\cal L}=\int dt \biggl[ f(\phi )\dot \phi^2
-if(\phi )\lambda_a\dot\lambda_a
-k\phi -kA_0\biggr]}
The last term here is a $d=1$ Chern-Simons term.  For gauge invariance
under large gauge transformations the coefficient $k$ must be an
integer.  In canonical quantization we pick the gauge $A_0=0$ and impose
Gauss law $G={\delta {\cal L} \over \delta A_0} =0$.  The Chern-Simons
term contributes a linear shift proportional to $k$ to the electric
charge $G$.  It corresponds to a background electric charge $k$.  Note
that in the system based on \coulLag, there are no charged fields and
therefore there is no solution to this Gauss law constraint.  Also,
because of the linear potential the energy is not bounded from below.
These issues will be addressed further below.

Naively this theory does not appear supersymmetric.  
The numbers of bosonic and fermionic fields do not agree.
Also, there is a linear potential in $\phi$, which is typically
forbidden in theories with 8 supercharges (such as $N=2$ 
theories in $4d$ and the more closely related $(0,8)$ 
models in $2d$).  In fact the Lagrangian \coulLag\ is invariant under
the following supersymmetry transformations:
\eqn\susy{\eqalign{
&\delta A_0=-i\epsilon_a\lambda_a\cr
&\delta\phi=i\epsilon_a\lambda_a\cr
&\delta\lambda_a=\epsilon_a\dot\phi-i{f^\prime(\phi) \over 2f(\phi)}
\epsilon_b\lambda_b\lambda_a.\cr}}
It is easy to check that the supersymmetry algebra closes (using the
equations of motion).  Thus, SUSY puts no constraint on the metric
function $f(\phi )$.

The linear potential manages
to be supersymmetric due to the Chern-Simons term, which
can only appear at one loop order.  (In the type $I^\prime$ 
quantum mechanics we will see that such a term indeed
appears out on a flat direction.)
The kinetic terms depend in an arbitrary fashion
on the scalar $\phi$ (in particular they are not protected
{}from loop corrections).  These features are reminiscent of 
$N=1$ supersymmetry in $4d$. 

Let us now consider adding $q$
fermionic multiplets charged under the $U(1)$. Each
contains a complex fermion $\chi_r,~r=1,\dots,q$.  
The Lagrangian now contains
\eqn\Lchione{{\cal L}_\chi=-i\bar\chi_r\dot\chi_r
-\bar\chi_r\phi\chi_r-\bar\chi_rA_0\chi_r-\sum_{r=1}^qm_r\bar\chi_r\chi_r.}  
Here we have included bare masses $m_r$ as well as the minimal
coupling to the gauge multiplet.  The $\chi s$ do not transform
under supersymmetry, and these terms preserve the supersymmetry
of the Lagrangian due to the cancellation between the
supersymmetry variations of $\phi$ and $A_0$  in \susy \ks.
Notice that integrating out the $\chi$s gives contributions
to the linear potential in \coulLag.    

How does this all work in the Hamiltonian formulation?  The
key fact is that the supercharges anticommute to the Hamiltonian
only up to gauge transformations.  Explicitly, 

\eqn\qalg{\{Q_a,Q_b\}=\biggl(H+\phi G+Z\biggr)\delta_{ab}}
where $G$ is the gauge constraint obtained by varying the Lagrangian
with respect to $A_0$, and $Z$ is a central term.  The Chern-Simons
term in \coulLag\ induces a shift by $-k\phi$ in the $\phi G$ term.
The linear potential term in \coulLag\ introduces a term $k\phi$ in $H$.
These cancel out in $\{Q_a,Q_b\}$.  A similar calculation explains
the terms involving the $\chi s$.  The gauge contribution
$\phi G$ to \qalg\ contains a term $-\bar\chi_r\phi\chi_r$.  
This cancels the term $\bar\chi_r\phi\chi_r$ in $H$.  
Let us now consider the mass terms for the $\chi s$, the terms
$\bar\chi_rm_r\chi_r$.  These appear in the Hamiltonian.  In order
to preserve supersymmetry, they must also appear with a negative
sign in the central charge $Z$.   
So although
the $\chi s$ do not appear in the supercharges, they appear in
the Hamiltonian, gauge constraint, and central charge in such
a way as to cancel in $\{Q_a,Q_b\}$.

There is a simple way to integrate out the fermions.  Their Lagrangian
is of the form $\bar \chi (i\partial_t - g V) \chi$ where $gV$ includes
the mass term, $\phi$ and $A_0$.  Integrating out $\chi$ leads to $\log
\det (i\partial_t - gV) $. Differentiating with respect to $g$ this is
$\int dt G(t,t)V(t)$ where $ 
G(t,s)= \exp i\int_s^t gV(u)du \theta (t-s)$ is the Green's function of
$ i\partial_t - g V$.  Using $\theta(0)=\ha$, we learn that the
Lagrangian resulting from integrating out the fermions is $\ha g V$.

This factor of $\ha$ here is significant.  It shows that our quantum
mechanical system with a single $\chi$ leads to a Chern-Simons term with
half integral coefficients.  Equivalently, the fermion determinant is
not gauge invariant under large gauge transformations.  These change the
sign of the fermion determinant leading to a global anomaly
\foot{For a similar global anomaly in quantum mechanics see, e.g.
\ref\efrs{S. Elitzur, Y. Frishman, E. Rabinovici and A. Schwimmer,
Nucl. Phys. {\bf B273} (1986) 93.}.}
This problem can be fixed by adding a bare Chern-Simons term with half
integral coefficient.

More explicitly, we consider a $U(1)$ gauge theory with $q$ fermions
with masses $m_r$ and a bare term $k(\phi+A_0)$.  Superficially, we need
to impose $k \in {\bf Z}$.  However, integrating out the fermions we
find in the effective action the terms
\eqn\newterms{-k(\phi+A_0) + {q \over 2} {\rm sign}(\phi-m_r) (\phi-m_r+A_0).}
Therefore, for consistency we need
\eqn\consis{k+q/2 \in {\bf Z}.}
In other words, the theory with odd $q$ without a bare potential is
inconsistent -- not gauge invariant.  This anomaly can be cancelled by
adding a bare term $k(\phi+A_0)$ with $k+1/2\in {\bf Z}$, which is also not
gauge invariant.

It is easy to understand this anomaly from a canonical quantization
perspective.  Consider $2q$ free real fermions.  The global symmetry
is $spin(2q)$.  The Hilbert space is $2^{q}$ dimensional in the two
spinors of $spin(2q)$.  Decomposing this under $ SU(q) \times U(1)$
we find integral $ U(1)$ charges for even $q$ and half integral
$U(1)$ charges for odd $q$.  Now we can gauge the $U(1)$ factor.  This
amounts to imposing Gauss law projecting on $U(1)$ invariant states.  If
$q$ is odd there is no such state.  If $q$ is even there are such
states.  We can also add the Chern-Simons term with coefficient $k$.
Now, the $U(1)$ charge is shifted by $k$.  Therefore, to have states in
the Hilbert space we need $ k+ q/2 \in {\bf Z}$.

\newsec{Applications to Type I' Theory}

\subsec{The Quantum Mechanical System}

Let us now consider the larger D0-brane system from which \coulLag\ is
obtained.  This consists of $n$ D0-branes propagating on
the interval $S^1/{\bf Z}_2$.  There
are also 16 D8-branes in the interval; let us call their positions $m_r$.
In \ulf\ the fields and tree-level Lagrangian for the $Spin(8)$ $N=8$  
theory which occurs in this type $I^\prime$ context were determined.
It has gauge group $Spin(n)$.  One finds   
quantum-mechanical coordinates
$$A^{IJ}_{0,9},~~~X^{IJ}_{1,...,8},~~~{\rm and}~~~x_{1,...,8}$$
where the $A^{IJ}$ are antisymmetric in the $Spin(n)$ indices $I$ and $J$
while the $X^{IJ}$ are in traceless symmetric representations and
the $x$s are singlets.   The fermionic superpartners are
$$\lambda^{IJ}_{a},~~~\theta^{IJ}_{\dot a},~~~{\rm and}~~~s_{\dot a}$$ 
with $\lambda_{a}$ in the adjoint, $\theta_{\dot a}$ 
in the traceless symmetric,
and $s_{\dot a}$ a singlet.  Here $a$ is an index in the ${\bf 8}_s$
of $Spin(8)$ and $\dot a$ is an index in the ${\bf 8}_c$.

This system, like the theory \coulLag, has different
numbers of bosons and fermions \ulf.  
Let us take $n$ even and calculate the vacuum energy out on the
classical flat 
direction in which $Spin(n)$ is broken to $U(n/2)$.  This
describes a cloud of zero branes away from the orientifold
plane (we are ignoring the D8-branes for now).  The
$U(1)$ factor in $U(n/2)$ encodes the motion
of the zero branes away from the orientifold plane.  
Let us denote by $\phi$ the scalar in this multiplet
(it is a component of $A_9$).  
The vacuum energy generated at one loop (i.e. the zero-point energy) is
\eqn\vacen{\Lambda_{0}=8n\phi}
The dynamics of the $U(1)$ multiplet is effectively
given by a Lagrangian of the form \coulLag.

Let us now consider the effects of the 16 D8 branes.
This introduces degrees of freedom coming from
the 0-8 strings.  Analysis of the worldsheet theory of
the 0-8 strings as in \S4.2\ of \joenew\ reveals that
the Neveu-Schwarz sector has vacuum energy $+{1\over 2}$,
so the lightest states are the Ramond-sector states.
Therefore, we must include fermions in the fundamental of $Spin(n)$
$$\chi^I_{r},~~\bar\chi^I_r ~~r=1,\dots,16.$$
Their Lagrangian contains terms involving the  $\chi$s
\eqn\Lchi{{\cal L}_\chi=-i\bar\chi_r \dot \chi_r-\bar\chi_rA_9\chi_r-
\bar\chi_r A_0\chi_r -m_r\bar\chi_r\chi_r.}
where the bare masses $m_r$ describe the
distance of the D0-branes from the r{\it th} D8-brane.

These fields contribute to the zero-point energy as well.  If 8
D8-branes sit at one orientifold ($m_r=0$), then the $\chi s'$
contribution to the vacuum energy is
\eqn\vacchi{\Lambda_\chi=-8n\phi,}
cancelling \vacen.  Thus, eight complex or equivalently sixteen real
$\chi$ fields are necessary to cancel the linear potential.
 More generally, the potential is piecewise linear
with singularities at $\phi=m_r$.  Clearly, the coefficient of the
Chern-Simons term is an integer and cannot be renormalized except at one
loop.  Since it is related by supersymmetry to the potential, the latter
also cannot be renormalized.

It is important to stress that this potential does not lead to explicit
supersymmetry breaking.  The effective Lagrangian with the potential
included is supersymmetric and supersymmetry is only spontaneously
broken.  This is achieved by adding the Chern-Simons term as in
\coulLag. 

\nref\edjoe{J. Polchinski and E. Witten, Nucl. Phys. {\bf B460} (1996) 525,
hep-th/9510169.}%

This piecewise linear function is the same one which appeared in the
space time analysis of \refs{\edjoe,\joenotes, \joenew} and in the four
brane discussion in \seiberg.  In space time it appeared at closed
string tree level.  On the four brane probe it renormalized the metric
at one loop.  This is one loop of open strings which is the same as a
closed string tree diagram \dkps.  In our problem it appears as a
potential for the zero branes.  It is amusing that the same function
affects different terms in the two different probes: the metric in the
four brane probe and the potential in the zero brane probe.  The
similarity between them is that these two terms are related by
supersymmetry to a Chern-Simons term.  Also, these two terms are
controlled by a nonrenormalization theorem and therefore this piecewise
linear function is exact.

\nref\redlich{A. N. Redlich, \prl{52}{1984}{18};
\physrev{29}{1984}{2366}.}%
\nref\agw{L. Alvarez Gaumez and E. Witten, \np{234}{1983}{269}.} %
\nref\niesem{A.J. Niemi and G.W. Semenoff, ``Axial-Anomaly-Induced
Fermion Fractionization and Effective Gauge-Theory Actions in
Odd-Dimensional Space-Times,'' Phys. Rev. Lett. {\bf 51} (1983) 2077.}%
\nref\ims{K. Intriligator, D. Morrison and N. Seiberg,
``Five-Dimensional Supersymmetric Gauge Theories and Degenerations of
Calabi--Yau Spaces,'' RU-96-99, IASSNS-HEP-96/112, hep-th/9702198.}%

This system is subject to the anomaly constraint \consis.   
A similar anomaly has already appeared in three dimensions in
\refs{\redlich, \agw} and in five dimensions in \refs{\niesem,\ims}.
In fact, the analysis in five dimensions is closely related to ours.
There the five dimensional theory occurs on a D4-brane probe in the
system of D8-branes.  Our analysis applies to a zero brane probe of the
same underlying spacetime physics.

\subsec{The Kinetic Terms}

As explained in \S2.1, the kinetic terms for $\phi$ and its fermionic
partners $\lambda_a$ are largely unconstrained by supersymmetry and
depend on an arbitrary function $f(\phi)$.  At tree level in the type
$I^\prime$ system we have $f_{tree}={1\over{\lambda_{I^\prime}}}$.  The
one-loop contribution is proportional to ${1\over\phi^3}$ by dimensional
analysis.  Direct computation yields a nonzero contribution.  This has
been previously computed\foot{We find that the $\chi$s do not contribute
to $f$.} in \ulf.  This gives an example of a situation where
dimensional reduction (in this case from the $(0,8)$ $2d$ theory discussed
above) does not preserve a nonrenormalization theorem.  In particular,
higher-loop contributions are not ruled out (and presumably contribute).
The loop expansion, which is an expansion in
$\lambda_{I^\prime}/\phi^3$, is sensible for large $\phi$.  But for
$\phi>{1\over{\sqrt{\alpha^\prime}}}$, we must also consider the effects
of the open string oscillator modes which exist in the type $I^\prime$
string theory.

The $\chi s$ do not run in the loops which contribute to this metric.
So even when 8 D8-branes are at either end of the interval, the metric
is nontrivial.  This is the configuration in which the spacetime dilaton
is constant.  As we will discuss more fully below, the constant dilaton
is reflected in the quantum mechanics by the fact that the contributions
\vacen\ and \vacchi\ to the potential cancel.  It is surprising that the
{\it metric} for $\phi$ and $X$ is not the same as the spacetime metric
found in \refs{\edjoe,\joenotes,\joenew}.  On the fourbrane probe
\seiberg, the metric is constant in this limit.  So we we see here a
case in which different probes have have qualitatively different
Lagrangians in the same background.

\subsec{Lorentz Invariance and the Supersymmetry Algebra}

The Hamiltonian 
on the $\phi$ branch of the $Spin(2)$ theory resulting from
the corrections computed above takes the form
\eqn\ham{H=\biggl({1\over\lambda_{I'}}
+{c\over\phi^3}+{\cal O}(\lambda_{I^\prime})\biggr)^{-1}P_\phi^2
+\biggl(16\phi-\sum_{r=1}^{16}|\phi-m_r|\biggr)
+{\lambda_{I^\prime}\over{2}}P_i^2+{1\over{\lambda_{I^\prime}}}.}
Here $P_\phi$ is the momentum conjugate to $\phi$, and
$P_i$ is the momentum conjugate to the center 
of mass position $x_i$ in the transverse 8 dimensions.

The last term is the bare mass of the D0-brane.
For large bare mass $M$, 
the Hamiltonian for a particle takes the form
\eqn\hamgen{H_0=M+{\vec P^2\over{2M}}.}
Our theory \ham\ satisfies this at the classical level
with $M={{1\over\lambda_{I'}}}$.  We must check that the constant piece
in the potential, as well as the linear term analyzed above,
is consistent with supersymmetry.
In this case the key is to keep track of
the central term $Z$ in the algebra \qalg.  
Although $Z$ appears indistinguishable from $H$ in
the algebra \qalg, it transforms as a singlet under
spacetime $Spin(8,1)$ Lorentz transformations.  The
term $M$ in $H$ is cancelled by a term
$-M$ in $Z$, leaving $\{Q_a,Q_b\}$ free of the constant piece.

The nontrivial metric generated by higher loop corrections
spoils the simple form \hamgen.  This is not a contradiction
with Lorentz invariance since the $S^1/{\bf Z}_2$ direction
breaks Lorentz invariance explicitly.  What is surprising is that
the nontrivial metric is generated even if 8 D8-branes
lie at each fixed point:  the nonvanishing graphs do not
involve the fields $\chi$.  This is the configuration for
which the dilaton tadpole in the spacetime theory cancels
locally \edjoe.      

For more generic D8-brane positions, the type I' dilaton varies over the
interval.  On a sublocus of the moduli space, which corresponds in the
dual heterotic string to the enhanced gauge symmetry locus in Narain
moduli space, the type I' dilaton blows up at the orientifold plane
($\phi=0$) \edjoe.  This suggests \joenotes\ that the enhanced gauge
bosons in the type I' theory are bound states of D0-branes localized at
$\phi=0$.  The D0-brane mass is ${1\over\lambda_{I'}}$, and this goes to
zero at $\phi=0$ on the enhanced symmetry locus.  This is consistent
with the linear potential we found within the quantum mechanics, which
causes the D0-brane to roll toward the orientifold plane.  As the
D0-brane approaches $\phi=0$, it becomes relativistic and the
approximation \hamgen\ breaks down.  In this regime, instead of starting
{}from \ham, one should presumably work with the Dirac-Born-Infeld action
to obtain the correct relativistic energy-momentum-mass relation
$E=\sqrt{\vec p^2+M^2}$.  In higher-dimensional probe theories
\refs{\seiberg,\morsei}, the enhanced symmetry locus in spacetime
corresponds to nontrivial renormalization-group fixed points on the
probe at the orientifold plane.  The analogue of that statement in the
$d=1$ case is the presence of a BPS state with gauge boson quantum
numbers corresponding to the global symmetry on the probe.  It would be
very interesting to make this explicit in the quantum mechanics, though
the importance of pair creation in the relativistic regime may render
this difficult.

\newsec{Applications to the Matrix Model}

One interesting issue in the proposed matrix model
formulation of M theory 
\ref\bfss{T. Banks, W. Fischler, S. Shenker, and L. Susskind,
``M Theory as a Matrix Model:  A Conjecture'', hep-th/9610043.}
is the origin of spacetime
gauge bosons.  One approach (as in \S9\ of \bfss ) is to
extrapolate BPS configurations in a given compactification
of the type IIA theory to
strong coupling,  relying on nonrenormalization theorems and time
dilation in the infinite momentum frame to argue that higher order
corrections to the weak coupling physics are negligible.  
This was applied to the $S^1/{\bf Z}_2$ case
with all the D8-branes at the ends of the interval
in \ks.  One finds using S and T dualities that the gauge
bosons arise from bound states of D0-branes with
the D8-branes at the
ends of the interval.  In particular, bound states of
odd numbers of D0-branes fall in the ${\bf 128}$ (spinor) of the $Spin(16)$
living on the D8-branes, while bound states of even D0-brane
number fall in the ${\bf 120}$ of $Spin(16)$.  The quantum mechanical
analysis of the bound states was extended in \lowe, where for
finite radius the role of winding open strings was explained.

This appears consistent as far as the states go, but our results here
suggest difficulties with the more general extrapolation to strong
coupling.  In the $N=16$ supersymmetric quantum mechanics describing
D0-branes in the type IIA theory, the metric is constrained to be flat
by supersymmetry, and there is potentially a nonrenormalization theorem
for the $({velocity})^4$ term describing scattering \refs{\dkps,\bfss}.
We see from the supersymmetry structure explained in \S2 -\S4\ that the
D0-brane system in type IIA on $S^1/{\bf Z}_2$ is already unconstrained
at the level of the metric or $({velocity})^2$.  (Recently,
renormalization of $({velocity})^4$ in a system with 8 supercharges was
observed in
\ref\dos{M.R. Douglas, H. Ooguri and S.H. Shenker, ``Issues in (M)atrix
Model Compactification,'' hep-th/9702203.}.)
In the matrix model context, the $N=8$ quantum mechanics we have studied
is supposed to represent a Horava-Witten domain wall in eleven
dimensional spacetime.  This is only possible when sixteen real $\chi$
fields are included in the Lagrangian, since otherwise there is a linear
potential due to the wall which spoils locality and translation and
rotation invariance far from it.  However, we have shown that even in
this case there is a nontranslationally invariant correction to the
metric in the effective Lagrangian far from the wall.  In each order of
perturbation theory this metric falls off with distance, but it is easy
to see that higher powers of inverse distance are accompanied by higher
powers of $N$.  Thus, in the large $N$ limit, it is only for $r \gg
N^{1/3} l_{11}$ that we can be confident that the effects of the wall
are negligible.

This is precisely analogous to the discussion of the coefficient of the
$(velocity)^4 $ interaction in \bfss\ .  There, a nonrenormalization
theorem was conjectured to resolve the problem, but in the present
context this cannot be the resolution.  Thus, if the $Spin(N)$ matrix
quantum mechanics is to represent M theory in the presence of a domain
wall, there must be some other principle (presumably valid only at large
N) which guarantees that the effects of the wall fall off with distance.
This principle might also be valid in the original matrix model,
eliminating the necessity of invoking an unproven nonrenormalization theorem.
We must also admit the possibility that these arguments show that the
matrix models do not give a correct description of eleven dimensional physics.

The fact that different probes of the same spacetime have different
moduli space metrics should not disturb us in general.  
The dilaton and metric may couple differently to different
probes.   There can even be spacetime curvature terms which appear in
the moduli space metric of some probes 
\ref\bachas{C.Bachas, private communication. }.  
However, it is somewhat confusing that in the case where the spacetime
equations of motion lead us to expect a flat space with constant
dilaton, zero branes and fourbranes seem to exhibit different physics.
The nontrivial metric on the moduli space of zero branes does not seem
to have an analog in four brane physics.  Recall however that the
physics of the metric on the four brane world volume field theory maps
into a potential in the zero brane quantum mechanics.  Perhaps the
nontrivial metric of the zero branes maps similarly into some higher
dimension term in the four brane Lagrangian.  Then we would be able to
keep a notion of an underlying spacetime physics which is simply
perceived differently by different branes.

\bigskip

\centerline{\bf Acknowledgements}

This work was supported in part by DOE grant DE-FG02-96ER40559.  
We would like to thank C. Bachas, S. Kachru, and S. Shenker for
discussions.

\listrefs

\end